\documentclass[reprint, amsmath, amssymb, aps, pra]{revtex4-2}

\usepackage{subfigure}
\usepackage{amsmath}
\usepackage{physics}
\usepackage{enumitem}
\usepackage{graphicx}
\usepackage[colorlinks=true,allcolors=blue]{hyperref}
\usepackage{xcolor}

\begin{document}
	
\title{\texorpdfstring{$\mathcal{PT}$}{PT}-symmetric quantum Rabi model}

\author{Xilin Lu\textsuperscript{1,2,3}}

\author{Hui Li\textsuperscript{1}}

\author{Jia-Kai Shi\textsuperscript{1}}

\author{Li-Bao Fan\textsuperscript{1}}

\author{Vladimir Mangazeev\textsuperscript{3}}

\author{Zi-Min Li\textsuperscript{1}}
\email{zimin.li@csu.edu.cn}

\author{Murray T. Batchelor\textsuperscript{2}}

\affiliation{\textsuperscript{1}Institute of Quantum Physics, Hunan Key Laboratory of Nanophotonics and Devices, Hunan Key Laboratory of Super-Microstructure and Ultrafast Process, School of Physics, Central South University, Changsha 410083, China\\
\textsuperscript{2}Mathematical Science Institute, Australian National University, Canberra ACT 2601, Australia\\
\textsuperscript{3}Research School of Physics, Australian National University, Canberra ACT 2601, Australia
}

\date{\today}

\begin{abstract}
In this work, we explore the $\mathcal{PT}$-symmetric quantum Rabi model, which describes a $ \mathcal{PT} $-symmetric qubit coupled to a quantized light field. 
By employing the adiabatic approximation (AA), we are able to solve this model analytically in the parameter regime of interest and analyze various physical aspects.
We investigate the static and dynamic properties of the model, using both the AA and numerical diagonalization.
Our analysis reveals a multitude of exceptional points (EPs) that are closely connected with the exactly solvable points in the Hermitian counterpart of the model.
Intriguingly, these EPs vanish and revive depending on the light-matter coupling strength.  
Furthermore, we discuss the time evolution of physical observables under the non-Hermitian Hamiltonian.
Rich and exotic behaviors are observed in both strong and ultra-strong coupling regimes.
Our work extends the theory of $ \mathcal{PT} $ symmetry into the full quantum light-matter interaction regime and provides insights that can be readily enlarged to a broad class of quantum optical systems.
\end{abstract}

\maketitle

\section{Introduction} 

Quantum mechanics assumes Hamiltonians that describe physical systems to be Hermitian to ensure real energy eigenvalues and unitary time evolution. 
However, there has been a recent surge of interest in non-Hermitian systems \cite{Moiseyev2014, Ashida2020}, especially those with parity-time ($\mathcal{PT}$) symmetry that manifest a transition from purely real to complex conjugate spectra \cite{Bender_2005}. 
$\mathcal{PT}$ symmetry not only is of fundamental importance but also has been applied across myriad fields \cite{ElGanainy2018, Oezdemir2019, Miri_2019}.  

The $\mathcal{PT}$-symmetric qubit, a two-level system with balanced gain and loss, is a paradigmatic model to demonstrate the $\mathcal{PT}$ symmetry \cite{Dogra_2021}. 
This simple solvable model effectively reveals the essential consequences of $\mathcal{PT}$ symmetry and has been realized across multiple platforms \cite{Naghiloo2019, Wang2021}.
Further research has explored the interaction between the $\mathcal{PT}$-symmetric qubit and classical light fields \cite{Joglekar_2014, Lee_2015, Xie2018}, with detailed analyses of energy spectra and phase transitions in these semi-classical models. 
However, $\mathcal{PT}$ symmetry in pure quantum systems, particularly in the context of light-matter interactions, has rarely been visited. 

The quantum interaction between light and matter is typically characterized by the quantum Rabi model (QRM), which describes a two-level system coupled to a single-mode quantized light field \cite{Rabi_1936, Braak_2011, Xie_2017}. 
Despite its simplicity, the QRM exhibits rich physics and has found applications ranging from quantum optics \cite{Fox2006} and condensed matter physics \cite{Irish_2007}, to molecular physics and state-of-the-art superconducting circuit quantum electrodynamics \cite{Forn_D_az_2010, Yoshihara_2016, Yoshihara_2018}. 

Motivated by the QRM, here we generalize the theory of $\mathcal{PT}$ symmetry into full quantum light-matter interaction systems. 
We begin by constructing a $\mathcal{PT}$-symmetric Hamiltonian from a realistic physical system and verifying its symmetry. 
Next, we interpret the model in the displaced oscillator picture and solve it analytically using the adiabatic approximation \cite{Crisp_1992, Irish_2005, Li2021GAA}.
We then investigate the static properties, including energy eigenvalues and eigenstates, both analytically and numerically. 
Interestingly, we observe an infinite number of exceptional points (EPs), which vanish and revive depending on the coupling strength. 
Additionally, we study the dynamics governed by the non-Hermitian Hamiltonian, with a particular focus on the time evolution of the mean photon number in the cavity and the qubit population.

\section{Model Hamiltonian}

\begin{figure}
	\centering
	\includegraphics[width=.9\linewidth]{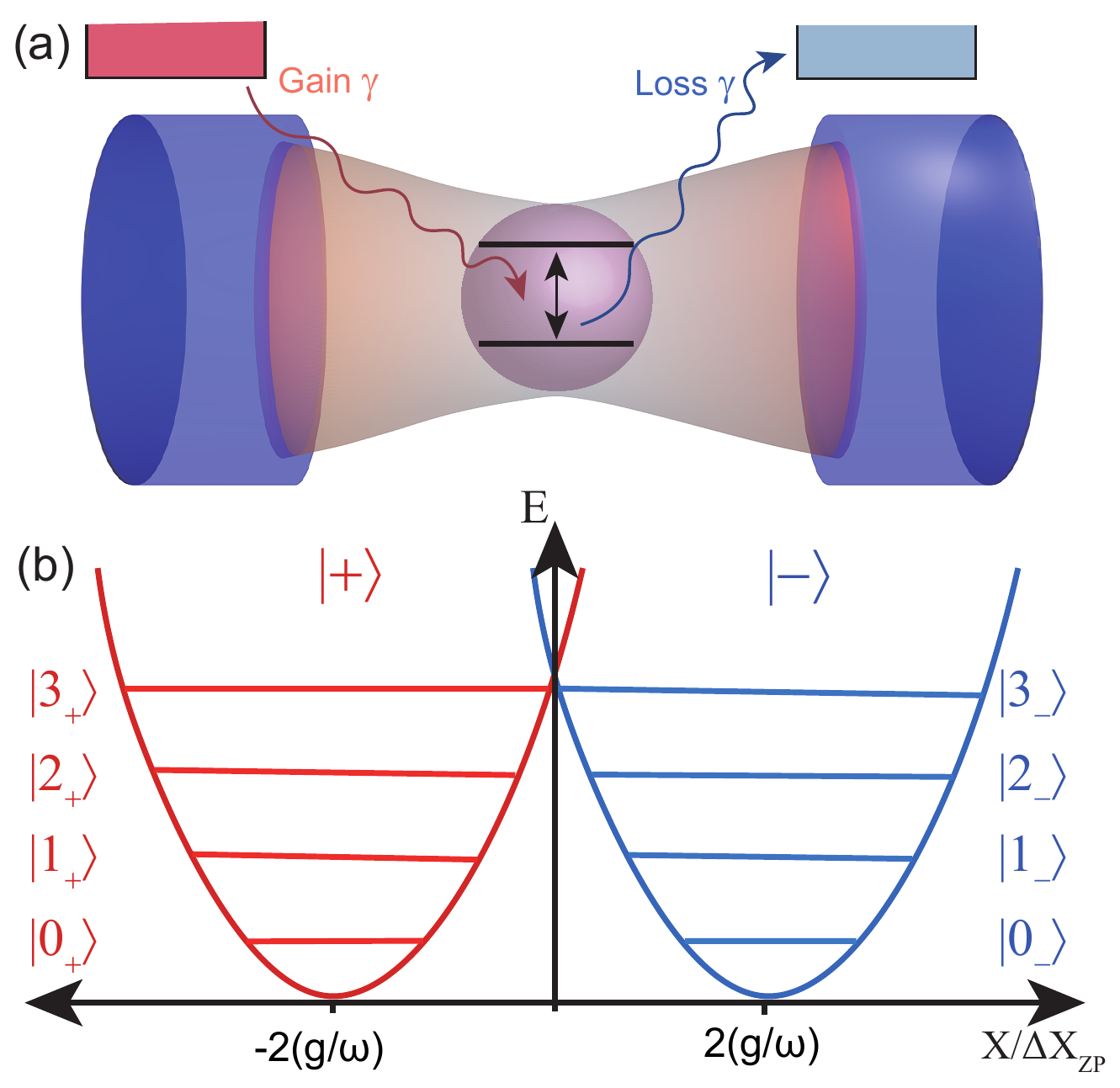}
	\caption{(a) Schematic of the system consisting of a $\mathcal{PT}$-symmetric qubit coupled to a cavity. (b) Two displaced oscillators, one with gain and the other with balanced loss, are coupled by the qubit term $\frac{\Delta}{2}\sigma_z$.}
	\label{schematic}
\end{figure}

The system we propose to study the $\mathcal{PT}$ symmetry is depicted in Fig.~\ref{schematic}(a), in which a non-Hermitian qubit is coupled to a single-mode cavity. 
This hybrid system can be described by the Hamiltonian
\begin{equation}\label{PTQRM}
 H = \dfrac{\Delta}{2}\sigma_z + \dfrac{i\epsilon}{2}\sigma_x +\omega a^\dagger a +  g\sigma_x \left(a^\dagger + a\right) ,
\end{equation}
where $\sigma_{x,z}$ are the spin 1/2 Pauli matrices, with $\Delta$ being the qubit level splitting and $\epsilon$ term being the driving or coupling between the two qubit states. 
The cavity mode with frequency $\omega$ is governed by the bosonic creation and annihilation operators $a^\dagger$ and $a$.
The coupling strength between the qubit and the cavity is denoted by $g$. 
Without loss of generality, all system parameters in this work are assumed to be non-negative. 

We begin our analysis with the $\mathcal{PT}$-symmetric qubit
\begin{equation}\label{Hqubit}
	H_q^z = \dfrac{\Delta}{2}\sigma_z + \dfrac{i\epsilon}{2}\sigma_x,
\end{equation}
for which the parity operator is $ \sigma_z$ and the time reversal operator is the complex conjugate operator.
The eigenvalues are simply $ \pm\sqrt{\Delta^2-\epsilon^2} $, indicating an exceptional point (EP) at $\epsilon/\Delta=1$, where the two eigenvectors coalesce and the eigenvalues become degenerate. 
When $\epsilon/\Delta$ crosses the EP, the eigenvalues change from real to complex, signaling the phase transition from $\mathcal{PT}$-symmetric to $\mathcal{PT}$-broken. 
More details on the \(\mathcal{PT}\)-symmetric qubit are given in the Appendix \ref{AppendixPT}. 

Regarding the light field, the effects of the $ \mathcal{PT} $ operators are equivalent to those of the bosonic parity transformation, leading to $a\rightarrow -a$ and $a^\dagger \rightarrow -a^\dagger$. 

Therefore, the combined parity operator for the hybrid quantum optical system (\ref{PTQRM}) is $\mathcal{P} = \sigma_z e^{i\pi a^\dagger a}$, which may be interpreted as the parity of the total excitation number in the system \cite{Li2021a}. 
Meanwhile, the time-reversal operation is still to take the complex conjugation. 
The symmetry of the full Hamiltonian can therefore be readily verified as
\begin{equation}
    (\mathcal{PT})^\dagger H \mathcal{PT} = H. 
\end{equation}
We thus refer to this system as the $\mathcal{PT}$-symmetric quantum Rabi model (PTQRM). 
Unlike previous approaches that change the coupling parameter $ g $ of the semi-classical Rabi model into an imaginary value \cite{Joglekar_2014, Lee_2015, Xie2018}, here we have taken a more physically intuitive route to obtain the pure quantum model by coupling a $\mathcal{PT}$-symmetric qubit to a quantized field. 

The PTQRM reduces to the well-known asymmetric quantum Rabi model (AQRM) if the imaginary term is replaced by the real version $\frac{\epsilon}{2}\sigma_x$ \cite{Braak_2011, Chen2012, Zhong_2014, Li_2015, Li_2016}. 
It reduces further to the standard QRM if $\epsilon=0$.

\section{Adiabatic approximation}

The non-Hermitian nature of the system, as described by Eq.~(\ref{PTQRM}), arises from the nonzero value of $\epsilon$.
Furthermore, the interplay between $\epsilon$ and $\Delta$ underpins our understanding of non-Hermitian physics \cite{ElGanainy2018, Dogra_2021}.
Consequently, neither $\Delta/\omega$ nor $\epsilon/\omega$ need to have large absolute values.
Under these conditions, the criteria for the adiabatic approximation (AA), given by $\Delta/\omega<1$, remains applicable \cite{Irish_2005, Li2021GAA}.
Thus, the physical phenomena of PTQRM can be explored through the analytical framework provided by the AA.
{It is worth noting that when the system deviates from the AA conditions, specifically when $\Delta/\omega \gtrsim 1$, the observed physics remains largely similar. However, in this regime, analytical treatments become challenging.}

The displaced oscillator picture offers a particularly intuitive understanding of the AA. 
In this representation, light-matter interactions manifest as qubit-dependent displacements in the light field \cite{Crisp_1992}. 
Such a representation becomes feasible by considering the degenerate qubit limit, where $\Delta = 0$.
In this case, the Hamiltonian becomes
\begin{equation}
    H_\text{do} = \omega a^\dagger a + g\sigma_x (a^\dagger + a) + \frac{i\epsilon}{2}\sigma_x .
\end{equation}
With respect to the basis of $\ket{\pm x}$, where $\sigma_x \ket{\pm x } =\pm \ket{\pm x}$, the Hamiltonian can be rewritten in the form of displaced creation and annihilation operators, namely
\begin{equation}\label{Hdo}
    H_\text{do}^\pm = \omega \left[\left(a^\dagger \pm \frac{g}{\omega}\right)\left(a \pm \frac{g}{\omega}\right)\right] \pm \frac{i\epsilon}{2} - \frac{g^2}{\omega}.
\end{equation}
It follows that $H_\text{do}$ is diagonalized by a unitary transformation generated by the position displacement operator $\mathcal{D}(\alpha) = \exp[-\alpha (a^\dagger -a)]$, with the displacement amplitude $\alpha=\pm g/\omega$ associated with the qubit states. 
This leads to the solutions
\begin{align}\label{Edo}
    \psi^\text{do}_{N,\pm} &=\ket{n_\pm,\pm x} =\mathcal{D}(\pm g) \ket{n} \otimes \ket{\pm x}, \\
    E^\text{do}_{N,\pm} &= n\omega - \frac{g^2}{\omega} \pm \frac{i\epsilon}{2},
\end{align}
where $ \ket{n_\pm} = \mathcal{D}(\pm g/\omega)\ket{n} $ are the displaced Fock states, also known as the generalized {or extended} coherent states. 

At an intuitive level, Eq.(\ref{Hdo}) describes two displaced harmonic oscillators. 
These oscillators have a displacement amplitude of $ \pm g/\omega $ and a gain or loss rate of $i\epsilon/2$, corresponding to the qubit states $ \ket{\pm x} $, as depicted in Fig.\ref{schematic}(b). 
This framework, termed the ``displaced oscillator picture'', is a well-established tool in light-matter interaction studies.

Next, we express the QRM Hamiltonian in terms of the displaced oscillator eigenstates, denoted by Eq. (\ref{Edo}), and introduce the parameter $\Delta$.
The qubit term, $ \frac{\Delta}{2}\sigma_z $, facilitates the coupling of these eigenstates, a process that can be analogized as tunneling between the two displaced oscillators. 
Central to the AA is the assumption that only the tunneling processes between eigenstates sharing the same quantum number $ n $ are taken into account, effectively disregarding higher-order terms.
By doing so, the PTQRM Hamiltonian is decomposed into block-diagonal form, consisting of infinitely many $ 2\times 2 $ blocks 
\begin{equation}\label{HnAA}
	H_n = n\omega - \dfrac{g^2}{\omega} + \dfrac{1}{2} \begin{pmatrix}
		i\epsilon &  \Omega_n  \\
		\Omega_n & -i\epsilon
	\end{pmatrix} ,
\end{equation}
where
\begin{equation}\label{OmegaAA}
	\Omega_n = \Delta e^{-2g^2/\omega^2} L_n\left(\dfrac{4g^2}{\omega^2}\right) ,
\end{equation}
with $ L_n(x) $ being the $ n $th-order Laguerre polynomial. 
The $2\times2$ matrix presented in Eq. (\ref{HnAA}) resembles a $\mathcal{PT}$-symmetric qubit. 
Yet, it also encompasses a parameter-dependent driving, $\Omega_n$, that integrates the effects of the light-matter interaction.
It is important to highlight that the AA essentially effectuates a change of basis. Specifically, it shifts from the (bare) photon-qubit basis 
\begin{equation}\label{barestate}
\ket{n,\pm}=\ket{n}\otimes\ket{\pm z}
\end{equation}
as detailed in Eq.(\ref{PTQRM}), transitioning to the displaced oscillator basis $\ket{n_\pm,\pm x}$ as described in Eq.(\ref{Edo}).
Considering the qubit's degree of freedom, this basis transformation results in a $\pi/2$ rotation on the Bloch sphere, while maintaining the $\mathcal{PT}$ symmetry.
A more comprehensive discussion on the $\mathcal{PT}$ symmetry postrotation is given in Sect. \ref{SectionDiscussions}.

To ascertain the eigenvalues of the PTQRM, we diagonalize $ H_n$, resulting in the eigenvalues
\begin{equation}\label{AAEigenvalues}
		E_{n}^\pm = n\omega - \dfrac{g^2}{\omega}\pm \dfrac{\sqrt{\Omega_n^2-\epsilon^2} }{2}. 
\end{equation}
Owing to the transformation from the infinite-dimensional matrix $ H $ to the two-dimensional matrix $ H_n $, the energy levels can now be characterized by a multitude of level pairs, denoted by the eigenvalues $E_n^\pm$. 
In our analysis, we initiate from the zeroth level $ (n=0) $. 
The term $ \sqrt{\Omega_n^2 - \epsilon^2} $ reveals the non-Hermiticity of the $n$th level pair. 
Notably, exceptional points (EPs) emerge when $\epsilon$ equates to $|\Omega_n|$.

The eigenstates corresponding to these eigenvalues are
\begin{equation}\label{AAEigenstates}
            \ket{\psi_{n}^\pm}=
    \mathcal{N}_n^\pm \begin{pmatrix}
        \dfrac{i\epsilon\pm\sqrt{\Omega_n^2-\epsilon^2}}{\Omega_n} \\ 
            \vspace{-.5em}\\
        1
    \end{pmatrix},
\end{equation}
which are expressed with respect to the displaced oscillator basis, as outlined in Eq.~(\ref{Edo}). 
The normalization factors for these eigenstates can be deduced as
\begin{equation}\label{key}
    \mathcal{N}_n^\pm = 
    \begin{cases}
        \dfrac{1}{\sqrt{2}}, & \abs{\Omega_n}\ge\epsilon ,\\
        \dfrac{\Omega_n}{\sqrt{2\epsilon^2\pm2\epsilon\sqrt{\epsilon^2-\Omega_n^2}}}, & \abs{\Omega_n}<\epsilon . 
    \end{cases}
\end{equation}
When $\epsilon=0$, the derived eigensystem aligns with the AA of the standard QRM, as detailed in Ref.~\cite{Irish_2005}.

Utilizing the AA results, we can derive analytical expressions for various physical observables. For instance, consider the fidelity between the two eigenstates of the $n$th level pair:
\begin{equation}\label{AAfidelity}
    F_n^\text{AA} = \abs{\braket{\psi_n^+}{\psi_n^-}}^2 = \begin{cases}
        {\epsilon^2}/{\Omega_n^2}, & \abs{\Omega_n}\ge\epsilon, \\
        {\Omega_n^2}/{\epsilon^2}, & \abs{\Omega_n}<\epsilon.
    \end{cases}
\end{equation}
Additionally, the mean cavity photon number is described as
\begin{equation}\label{AAphoton}
    \expval{a^\dagger a}_n^\pm = n + g^2/\omega^2.
\end{equation}
The qubit expectation value is given analytically by
\begin{equation}\label{AAqubit}
    \expval{\sigma_z}_n^\pm = \begin{cases}
        \pm \dfrac{\sqrt{\Omega_n^2-\epsilon^2}}{\Omega_n}, & \abs{\Omega_n}\ge \epsilon, \\ 
        0, & \abs{\Omega_n}< \epsilon.
    \end{cases}
\end{equation}
Using this expectation value, the qubit population can be determined as
\begin{equation}\label{qubitpopulation}
    W_n^\pm = \frac{1}{2}\left(1+\expval{\sigma_z}_n^\pm\right).
\end{equation}

As depicted in Fig.~\ref{fidelity}, there is a clear agreement between the AA expressions and exact results.
This consistency provides deeper insights into the intricate dynamics of $\mathcal{PT}$-symmetric physics.

\section{Spectrum, exceptional points, and phase transitions}

\begin{figure*}[bt]
	\centering
	\includegraphics[width=\linewidth]{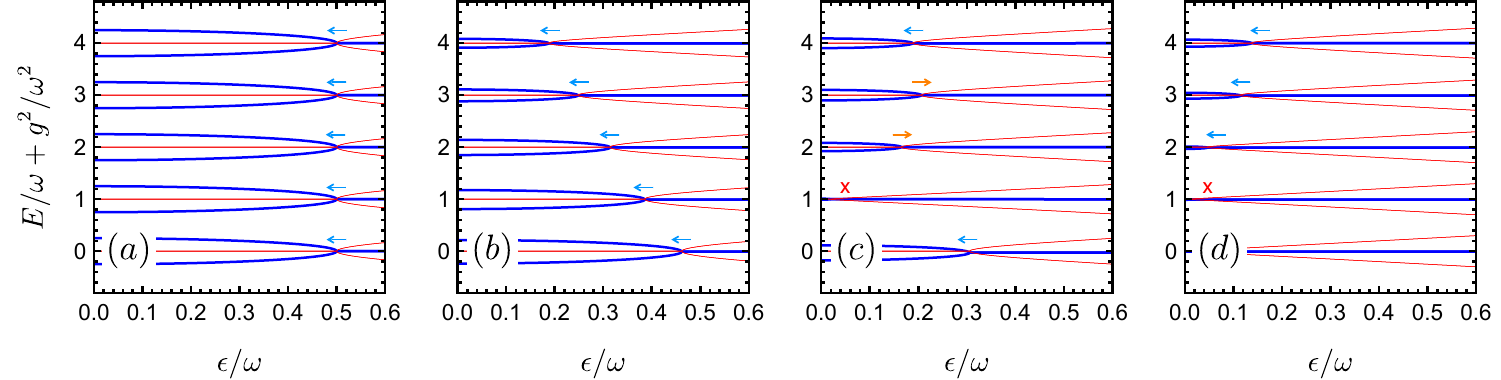}
	\caption{Vanishing and revival of the EPs. (a)-(d) Real (blue thick) and imaginary (red thin) energy spectra of the PTQRM with respect to the qubit parameter $\epsilon$, with the coupling strength $g/\omega = 0, 0.2, 0.5$, and $ 1.8 $, respectively. For each figure, $\Delta/\omega$ is set to 0.5. For clarity, imaginary parts of the eigenvalues have been rescaled by adding the corresponding level index $n$.
    The arrows in the figures denote the moving directions, blue for vanishing and orange for reviving, of the EPs if $g$ is to be increased. }
	\label{EepsilonSpectrum}
\end{figure*}

\begin{figure}[bt]
    \subfigure{\includegraphics[width=\linewidth]{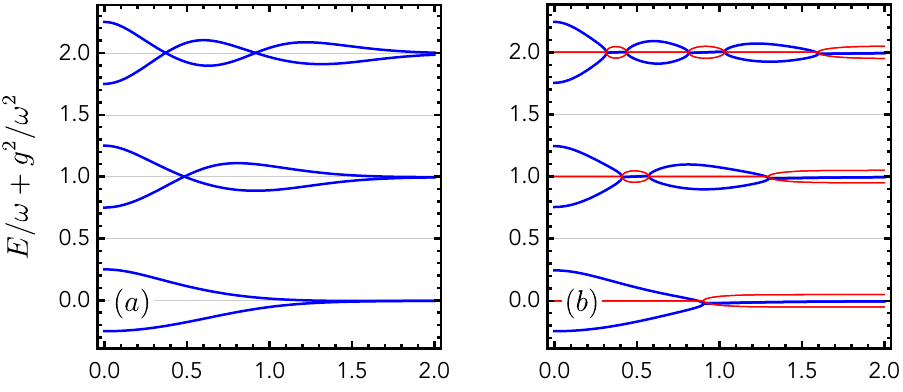}} \\ \vspace{-1em}
    \subfigure{\includegraphics[width=.98\linewidth]{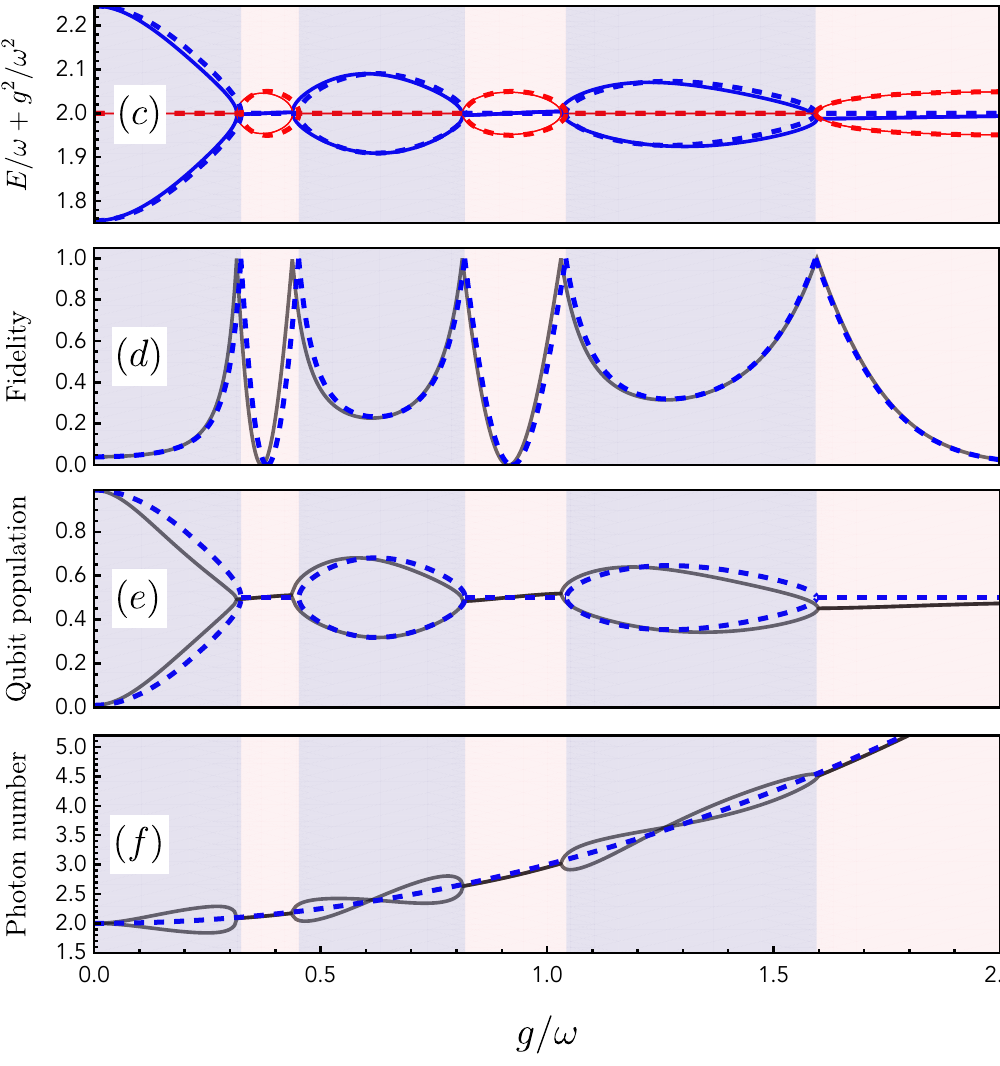}}
	\caption{
        The energy spectra of (a) the QRM and (b) the PTQRM with respect to the coupling strength $g/\omega$. The real and imaginary parts of the PTQRM eigenvalues are denoted with blue and red lines, respectively. 
        For the third pair levels, the plots are
		(c) the energy spectrum,
		(d) the fidelity, 
		(e) qubit populations, and 
		(f) mean photon numbers for the corresponding two eigenstates, against coupling strength $g/\omega$. 
		Relevant parameter values are $\Delta/\omega=0.5$ and $\epsilon/\omega=0.1$. Analytic results of AA are depicted in dashed lines. The system is in the $\mathcal{PT}$-symmetric phase in the blue (dark gray) regions and the $\mathcal{PT}$-broken phase in the red (light gray) regions. }
	\label{fidelity}
\end{figure}

In our subsequent analysis, we delve into the spectral structure and the novel non-Hermitian dynamics inherent in the PTQRM. 
We employ both numerical diagonalization and analytical insights from the AA for a more comprehensive understanding. 
Analyzing the spectrum, we discern numerous EPs, akin to those observed in the $\mathcal{PT}$-symmetric qubits.
Notably, these EPs exhibit additional modulation attributable to the cavity. 
Intriguingly, the cavity introduces an exotic phenomenon: the vanishing and revival of EPs. 

To observe the interesting EP phenomena under light-matter interaction, we tune the ratio of the qubit parameters $\Delta$ and $\epsilon$, as shown in Fig.~\ref{EepsilonSpectrum}. 
In the decoupled case, all level pairs behave the same as the spectrum of the $\mathcal{PT}$-symmetric qubit, only with the energy raised by the cavity eigenvalues. 
When the coupling strength $ g $ is increased, EPs are shifted due to the interaction, and their overall behavior goes through three stages: vanishing, oscillating revival, and vanishing again. 
It can be seen from Fig.~\ref{EepsilonSpectrum}(b) that EPs at higher levels move faster. 
Therefore, EPs from high to low vanish in sequence with the increase of $ g $.  
Owing to the light-matter interaction, the EPs may reappear in an oscillatory manner. 
Finally, when the interaction approaches a certain strength, all EPs tend to disappear in sequence, as illustrated in Fig.~\ref{EepsilonSpectrum}(d) with $ g/\omega=1.8 $. 

The phenomenon of EPs' vanishing and revival can be inferred from the PTQRM spectrum with respect to the coupling strength $ g $, as shown in Fig. \ref{fidelity}(b). 
It is noteworthy that the majority of EPs stem from the exactly solvable Juddian points, the level crossings, in the QRM \cite{Li_2015}, {whose spectrum is depicted in Fig.~\ref{fidelity}(a)}. 
The locations and numbers of Juddian points on each pair of energy levels are predicted by the defining polynomials \cite{Judd_1979, Kus1985, Li_2015, Wakayama_2017, Kimoto_2020}.
In the AQRM, where the $\epsilon$ term is real, the Juddian points manifest as diabolic points (DPs) in the three-dimensional parameter space \cite{Batchelor2015, Li_2021}. 
Each DP bifurcates into a pair of EPs when $\epsilon$ assumes an imaginary value.

In addition to the spectrum, the eigenstates also exhibit novel $\mathcal{PT}$-symmetric physics due to the light-matter interaction. 
To explore the eigenstates, we take the second-level pair, determined by $ H_2 $, as an example. 
The eigenspectrum together with the corresponding fidelity, qubit population, and mean photon number is shown in Fig.~\ref{fidelity}, with the AA results denoted by the dashed lines. 

Owing to the intricate modulation induced by the light-matter interaction, there is no standard overall real-to-complex phase transition when considering all energy levels \cite{Bender_2005}. 
For a given set of parameter values, there will always be imaginary eigenvalues when photon numbers reach sufficiently high values.
Nevertheless, it proves beneficial to delineate ``local'' phases for individual level pairs of the system's eigenstates. 
To be precise, for a specified level index $ n $, the system is considered to be in the $\mathcal{PT}$-symmetric (PTS) phase if the eigenvalues $ E_n^\pm $ are real, and in the $\mathcal{PT}$-broken (PTB) phase if they are complex. 
Such a definition becomes particularly pertinent when the cavity contains only a limited number of photons and the Hilbert space remains finite-dimensional.
These phases can be visualized in Fig.~\ref{fidelity}, distinguished by blue (PTS) and red (PTB) backgrounds.

In Hermitian systems, the eigenstates associated with different eigenvalues are orthogonal. 
Contrastingly, in non-Hermitian scenarios, eigenstates are typically non-orthogonal and even parallel at EPs. 
For two pure states, their fidelity is defined as $ F_{12}=\left|\braket{\psi_1}{\psi_2}\right|^2 $. 
Two states are orthogonal if $ F_\text{12}=0 $ and parallel if $ F_{12}=1 $. 
As illustrated in Fig.~\ref{fidelity}(d) that the two eigenstates are generally not orthogonal, with non-zero fidelity. 
Remarkably, eigenstates align perfectly at multiple EPs, with the fidelity being unity, a characteristic of EPs that does not exist in Hermitian systems. 
Another intriguing characteristic observed is that the eigenstates are orthogonal at some specific points between the EPs, which is absent in normal non-Hermitian systems. 
Interestingly, these orthogonal points coincide with the original Juddian points in the standard QRM \cite{Li_2015}, and their positions are independent of the value of $\epsilon$.  
The nature of these exotic orthogonal points can be understood from the perspective of the AA. 
In the AA for the standard QRM, Juddian points appear when the Laguerre polynomials in $\Omega_n$ vanish.
Consequently, the eigenvectors in Eq.~(\ref{AAEigenstates}) become $ \psi_n^+=(1,0)^T $ and $ \psi_n^-=(0,1)^T $, which are trivially orthogonal. 

Figure.~\ref{fidelity}(e) depicts the qubit population using dashed lines, while the analytical results from Eq.~(\ref{AAqubit}) are represented by solid lines. 
This behavior is similar to the $\mathcal{PT}$-symmetric qubit dynamics, as discussed in Ref.~\cite{Dogra_2021}.
Specifically, for a zero coupling strength, $g/\omega=0$, the qubit and cavity are decoupled, with the PTQRM eigenstates manifesting as the bare states $\ket{n,\pm}$ from Eq.~(\ref{AAEigenvalues}).
Here, the corresponding qubit populations are trivially 1 for $\ket{+}$ states and 0 for $\ket{-}$ states.
Upon introducing non-zero coupling, entanglement between the qubit and cavity states ensues, instigating a blend of the two qubit states and a resultant shift in qubit populations. 
The eigenstates coalesce at the first EP, aligning proximally to the Bloch sphere's equator, leading to equal qubit populations for both eigenstates.
In the subsequent PTB phase post this EP, the qubit states consistently traverse the Bloch sphere's equator, maintaining this uniform population distribution.
As \(g\) increases, the system passes through several PTS and PTB zones, all of which exhibit analogous behavior.

In Fig.~\ref{fidelity}(f), we present the variation of the cavity photon number as a function of the coupling strength.
The associated AA analytical model is captured by Eq.~(\ref{AAphoton}).
Consistent with expectations, the mean photon number within the cavity grows with increasing coupling strength, aligning with findings from Ref.~\cite{Rossatto_2017}. 
It is worth noting that discrepancies arise between the exact results and the AA photon number, given by $\expval{a^\dagger a}_n^\pm = n + g^2/\omega^2$, particularly within the PTS regions. This divergence can be attributed to the inherent assumptions within the AA framework. 
In the reduced subspace, the photon number is definite for a specific $ n $ and displacement $g$. 
Therefore, the deviations in Fig.~\ref{fidelity}(d) cannot be captured solely by the AA and require the inclusion of higher-order tunneling effects \cite{Li2021GAA}. 

~

\section{Dynamics of physical observables}

The dynamics of the PTQRM manifests rich physics absent in its Hermitian or semi-classical counterparts. 
To reveal this, we focus on the time evolution of the mean photon number {and qubit population} by numerically solving the time-dependent Schr\"odinger equation {(TDSE) across varied parameter domains}. 

\subsection{{Dynamics in the strong coupling regime}}

\begin{figure*}[tb]
    \includegraphics[width=\linewidth]{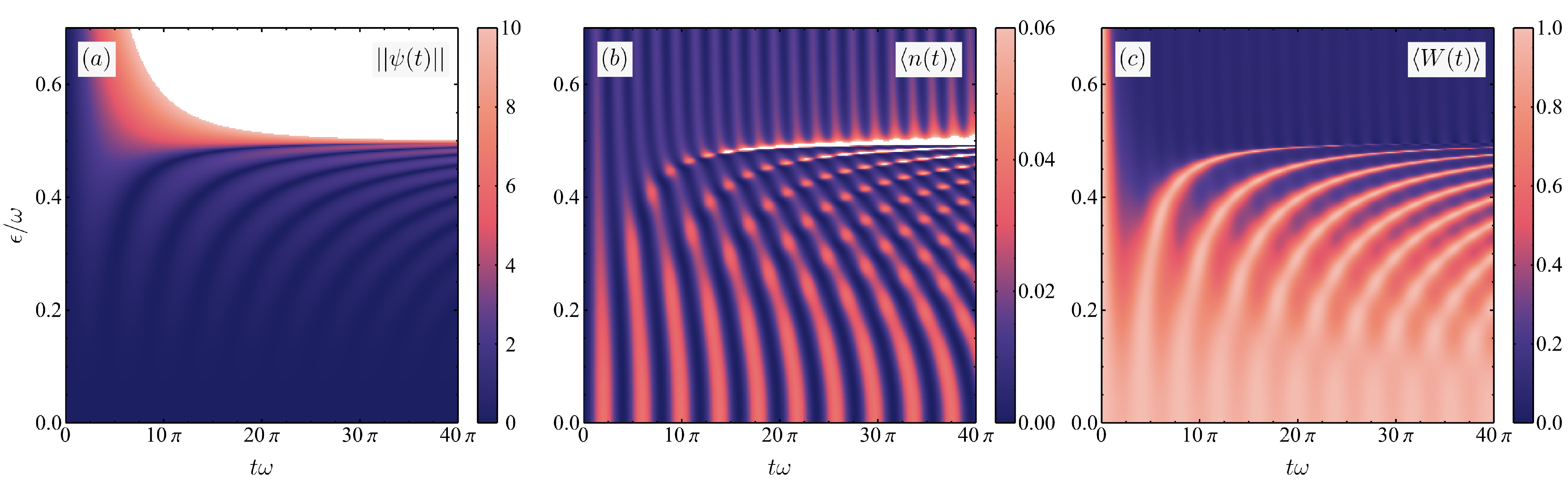}
    \caption{Time evolution of the (a) norm of the wave function (which is rescaled with logarithms and capped at 10), (b) cavity photon number, and (c) qubit population of the PTQRM in the strong coupling regime, with $g/\omega=0.05, \Delta/\omega = 0.5$ and the value of $\epsilon/\omega$ ranging from 0 to 0.7. The initial state is set as $\ket{0,+}$.}
    \label{SCfigs}
\end{figure*}

In our analysis, we begin by focusing on the strong coupling (SC) regime, wherein the light-matter interaction strength is substantially lower than both the qubit and photon frequencies, that is, \(g\ll \omega, \Delta\). 
Traditionally, the SC regime requires a coupling strength surpassing the system's decay rates \cite{Fox2006,Frisk_Kockum_2019}. 
However, within the context of the non-Hermitian PTQRM, we still identify this regime as SC even if the loss rate \(\epsilon\) could exceed the coupling strength \( g \). 
This classification arises from the fact that the gain rate \(\epsilon\) ensures sustained dynamics prior the exponential decay \cite{ElGanainy2018}. 

Within the Hermitian SC regime, the rotating wave approximation (RWA) is typically employed to the QRM, leading to the well-known Jaynes-Cummings model (JCM) \cite{Jaynes_1963, Shore_1993,Fox2006}. 
This approximation is valid, given that the effects of counter-rotating terms (CRTs) (those terms which do not conserve excitation numbers) are negligible in this regime. 
A key consequence of the RWA is that the otherwise infinite-dimensional Hilbert space is decomposed into infinitely many two-dimensional subspaces, thereby facilitating the analytic solvability \cite{Li2021GAA}. 
The SC regime's dynamics are primarily characterized by Rabi oscillations between various bare photon-qubit states, signaling the exchange of excitations between the cavity and the qubit.  

In this part, we explore the non-Hermitian dynamics in the SC regime. 
Contrasting with the Hermitian case, the non-Hermitian terms (gain and loss) couple all the basis states, rendering the Hilbert space undecomposable via the RWA. 
To carry out the calculations, we start from the initial bare photon-qubit state \(\ket{0,+}=\ket{0}\otimes\ket{+z}\), indicating an empty cavity and a qubit in its excited state. 
Subsequently, the non-Hermitian dynamics are examined across a range of $\epsilon$ values, while holding the other system parameters constant at \(\Delta/\omega=0.5\) and \(g/\omega=0.01\). 
Based on Fig.~\ref{EepsilonSpectrum}(a), the EPs for the lowest energy levels should lie proximately at \(\epsilon/\omega=0.5\).

We begin by investigating the wave function's norm within the PTQRM framework. 
As mentioned before, the norms of wave functions, defined by
\begin{equation}
    \norm{\psi(t)}=\sqrt{\braket{\psi(t)}{\psi(t)}},
\end{equation} 
deviate from conservation due to the Hamiltonian's non-Hermitian nature. 
However, it is worth noting that the expectation values of physical observables in non-Hermitian systems necessitate wave function norm renormalization, featuring the norm dynamics as a pivotal indicator of non-Hermitian effects.
The dynamics, after taking logarithms and capped at 10, are illustrated in Fig.~\ref{SCfigs}(a).
The figure distinctly shows that, in contrast to the Hermitian case where the norms are inherently conserved, the norm of the wave function oscillates in the PTS regime (\(\epsilon/\omega<0.5\)) and diverges in the PTB regime (\(\epsilon/\omega>0.5\)).
These behaviors are similar to the dynamics previously observed in the $\mathcal{PT}$-symmetric qubit \cite{Dogra_2021}.  

Proceeding further, we calculate the renormalized cavity photon number, represented as
\begin{equation}\label{photonnumber}
    \expval{n}(t) =\expval{a^\dagger a}(t) = \frac{\bra{\psi(t)}a^\dagger a \ket{\psi(t)}}{\braket{\psi(t)}{\psi(t)}},
\end{equation}
which holds significance in the realm of light-matter interaction analysis. 
Referring to Eq.~(\ref{photonnumber}), the associated dynamics, depicted in Fig.~\ref{SCfigs}(b), deviate noticeably from those in the Hermitian context.  
The figure reveals that photon numbers oscillate across both the PTS and PTB phases. 
Intriguingly, while the PTS regime sees the oscillation characteristics -- frequency and amplitude -- being influenced by the non-Hermitian term $\epsilon$, they stabilize and remain invariant in the PTB regime. 

The qubit population stands as a pivotal quantity in the domain of quantum information sciences.
We compute this quantity using Eq.~(\ref{qubitpopulation}), and thus its time evolution
\begin{equation}
    W(t) =  \frac{1}{2}\left(1+\bra{\psi(t)}{\sigma_z}\ket{\psi(t)}\right).
\end{equation}
Referring to the outcomes presented in Fig.~\ref{SCfigs}(c), it is evident that within the PTS regime, the quantity oscillates around the originating qubit state, $\ket{+z}$. This particular state corresponds to the north pole on the Bloch sphere. 
Transitioning to the PTB phase, the population gravitates towards the equatorial state on the Bloch sphere, registering a population value of 0.5.  
Notably, subtle oscillations remain manifest in this PTB phase. 

Through the dynamics of the three quantities demonstrated in Fig.~\ref{SCfigs}, the transition from the PTS phase to the PTB phase becomes evident, specifically around the point \(\epsilon/\omega=0.5\).
In detail: the norm shifts from oscillatory behavior to a divergent trajectory; both the photon number and qubit population evolve from oscillations dependent on \(\epsilon\) to oscillations that remain indifferent to variations in \(\epsilon\).
This transition can be further elucidated when considering a low-dimensional subsystem accommodating a maximum of 1 photon. 
Such a system can be described by the \(4\times 4\) Hamiltonian matrix
\begin{equation}
    H_\mathrm{low} = 
    \begin{pmatrix}
        -\frac{\Delta}{2} & \frac{i\epsilon}{2} & 0 & g\\ 
        \frac{i\epsilon}{2} & \frac{\Delta}{2} & g & 0 \\ 
        0 & g & \omega-\frac{\Delta}{2} & \frac{i\epsilon}{2} \\
        g & 0 & \frac{i\epsilon}{2} & \omega + \frac{\Delta}{2}
    \end{pmatrix},
\end{equation}
which is expressed with respect to the basis \(\{\ket{0,\pm},\ket{1,\pm}\}\). 
Within this matrix, coupling terms induce transitions across the basis states. 
It is noteworthy to mention that this system encapsulates both Hermitian and non-Hermitian transition mechanisms.
For the Hermitian processes driven by \(g\), the RWA can be applied, allowing analysis through the JCM. 
This typically results in the manifestation of Rabi oscillations. 
Conversely, the non-Hermitian mechanisms stemming from \(\epsilon\) terms bear similarities to the dynamics of the $\mathcal{PT}$-symmetric qubit: they oscillate based on the gain and loss rates in the PTS phase and stabilize in the PTB phase \cite{Dogra_2021}. 
In the PTS regime, the dynamics introduce an intriguing interplay, with non-Hermitian oscillations challenging the established JCM behavior. 
In the PTB phase, however, the non-Hermitian mechanisms quickly reach a steady state, leaving only the JCM dynamics active and culminating in oscillations that are indifferent to variations in \(\epsilon\).
Drawing a parallel, it is evident that the PTQRM dynamics in the SC regime share considerable similarities with those of the $\mathcal{PT}$-symmetric qubit \cite{Dogra_2021}. 
Nonetheless, PTQRM introduces a unique layer of complexity by fostering competition in non-Hermitian dynamics, primarily by enabling energy exchanges between light and matter.

In the SC regime, while the immediate impact of CRTs -- which couple distinct subspaces -- may seem insubstantial, their cumulative effects can be significant over time. 
Consequently, with prolonged duration, the probability may escape from the predominant subspace.
For sufficiently high coupling strengths, even over short timescales, this effect becomes non-negligible, leading to markedly distinct dynamic behaviors.

\subsection{{Dynamics in the ultrastrong coupling regime}}

\begin{figure}[tb]
    \centering
    \includegraphics[width=\linewidth]{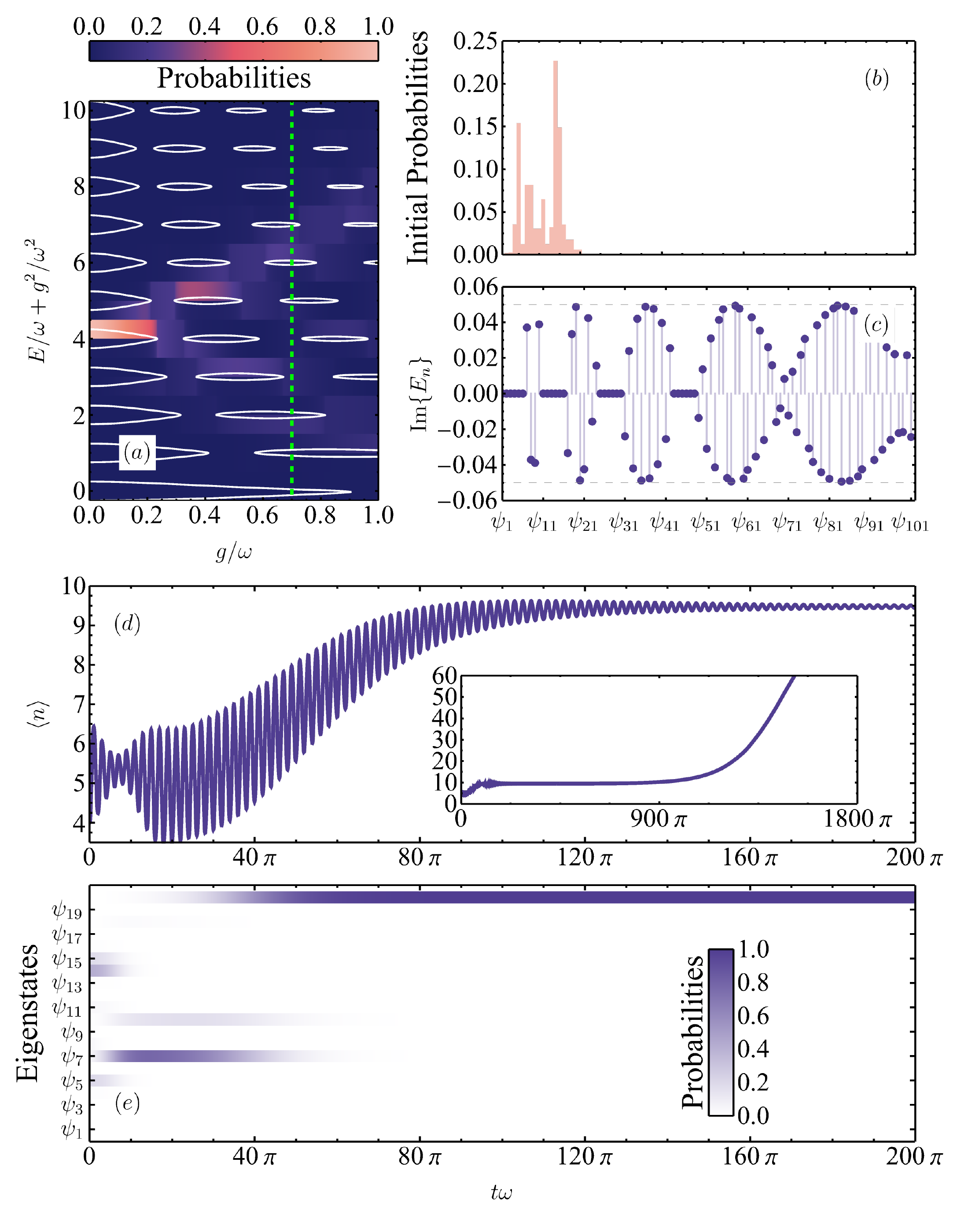}
     \caption{
        (a) The real spectrum of the PTQRM, with background colors indicating the mapping of the initial bare state $\ket{4,+}$ onto the eigenstates.
        The vertical dashed green line indicates the coupling strength $g/\omega=0.7$. 
        (b) Initial probabilities with respect to the basis of eigenstates with $g/\omega=0.7$, mapped from the initial bare state $\ket{4,+}$. 
        (c) Imaginary parts of eigenvalues for the lowest 101 eigenstates with $g/\omega=0.7$. The dashed grid lines denote the max and min values $\pm \epsilon/2$.
        Time evolution of (d) the cavity photon number and (e) eigenstate probabilities with $\ket{\psi(0)}=\ket{4,+}$. The inset in (d) is the long-term dynamics where all three stages are present. 
        Relevant parameter values are $\Delta/\omega=0.5$ and $\epsilon/\omega=0.1$.
        }
    \label{PhotonDynamicsFig}
\end{figure}

Ultrastrong coupling (USC) between light and matter is reached when the coupling strength \(g\) is comparable to the cavity or qubit frequencies \cite{Frisk_Kockum_2019, Forn_D_az_2019}. 
In the USC regime, CRTs significantly influence the dynamics, making the application of the RWA infeasible for decomposing the Hamiltonian into \(2\times 2\) matrices \cite{Irish_2005, Li2021GAA}.
Consequently, any wave function becomes a superposition of all eigenstates. 
Decomposing an arbitrary wave function with respect to the eigenstate basis reveals components both within the PTS and PTB regimes. 
This leads to a system evolution that intricately blends PTS and PTB dynamics in the USC regime, which is sharply different from the SC regime. 
Our subsequent analysis will delve into the time evolution of the cavity mean photon number, determined numerically via the TDSE, highlighting the profound effects of the ultrastrong light-matter interaction.

The long-term dynamics of the mean photon number {in the USC regime} can be classified into three distinct stages: firstly, it exhibits oscillations around the initial state; secondly, it converges towards eigenstates characterized by the locally largest imaginary eigenvalues; and finally, in the long term, it diverges towards infinite-photon states.
Notably, the duration of the first two stages is modulated by both system parameters and the chosen initial states.

In the following, we discuss in detail the case $g/\omega=0.7$ as an example. 
We start from the initial bare state \(\ket{4,+}=\ket{4}\otimes\ket{+z}\), which represents a system configuration with four photons in the cavity and the qubit in its excited state.
We can express this initial state in the context of the eigenstate basis.
To determine the component probability amplitudes \(p_n\) , we project the initial state onto these eigenstates, yielding 
\begin{equation}
    p_n = \abs{\braket{4,+}{\psi_n}}^2.
\end{equation} 
Here, eigenstates are denoted $\ket{\psi_n}$, with $n$ indexed in order of increasing eigenvalues.
For computational considerations, our analysis includes only the first 101 eigenstates.
It is worth noting that expanding the analysis to a broader set of eigenstates would still result in the observed three-stage behavior.

Figure \ref{PhotonDynamicsFig}(a) illustrates the real eigenvalues of the lowest 22 eigenstates of the PTQRM.  
The background colors represent the respective probabilities of these eigenstates in the initial configuration \(\ket{4,+}\). 
Tracing the line $g/\omega=0.7$ in Fig.~\ref{PhotonDynamicsFig}(a), we observe several pairs of eigenstates in the PTB phase.
Notably, the ninth pair exhibits the largest imaginary energies. 

To further elucidate, Fig.~\ref{PhotonDynamicsFig}(b) presents the probability distributions across these eigenstates.
While the majority of probabilities cluster around the lower eigenstates, we emphasize that none of the higher eigenstates have a strict probability of 0.
This phenomenon follows from above that every wave function manifests as a superposition of all the eigenstates in the USC regime. 
Although these probabilities in higher eigenstates may initially appear insignificant, their amplification by imaginary eigenvalues can propel them to prominence. 
Corroborating this, we plot the imaginary eigenvalues Im(\(E_n\)) of the lowest 101 eigenstates in Fig.~\ref{PhotonDynamicsFig}(c).
Notably, imaginary energies are more prominent in higher-level eigenstates, and their locally maximal imaginary eigenvalues are closer to the theoretical maximum of \(\epsilon/2\). 
The interplay between these probability distributions and imaginary eigenvalues gives rise to the distinct three-stage dynamics, which will be further explored.

In Fig.~\ref{PhotonDynamicsFig}(d), we present the evolution of the mean photon number, starting from the bare state $\ket{4,+}$. 
Initially, the system exhibits oscillations around its initial state, stemming from the interplay among dynamics of the high-probability eigenstate components.  
As the evolution progresses, the system stabilizes, with the mean photon number settling around 9. 
This plateau arises as the norms of the PTB states amplify and oscillatory behaviors diminish, notably post \(100\pi\). 
The stable behavior is sustained until the system eventually diverges towards infinite-photon states, which is evident from the long-term dynamics demonstrated in the inset.
In numerical calculations, however, the infinity is unattainable due to our Hilbert space truncation, and the maximal photon number caps at approximately 100.

The transition between the first two stages stands out as particularly interesting. 
In this process, the imaginary energies $\Im(E_n)$ dictate the growth rates of the corresponding eigenstates $\ket{\psi_n}$. 
Over time, an eigenstate with the largest imaginary energy prevails through exponential growth, overshadowing their initial probabilities. 
This phenomenon is exemplified when projecting the bare state $\ket{4,+}$ onto the eigenstates.
Remarkably, the dominant eigenstate $\ket{\psi_{20}}$, despite its initial probability being less than 0.01, quickly rises to prominence, as illustrated in Fig.~\ref{PhotonDynamicsFig}(e).

\section{Further discussions}\label{SectionDiscussions}

In the preceding sections, we have proposed the PTQRM Hamiltonian and analyzed its exotic properties in both statics and dynamics. 
Moving forward, we aim to provide a deeper understanding of the PTQRM Hamiltonian. 
To this end, we demonstrate an alternative representation of the PTQRM to frame the non-Hermitian Hamiltonian within tangible physical contexts.
We also juxtapose the non-Hermitian dynamics against the outcomes derived from the quantum master equation, which is the standard approach to deal with open systems. 
Lastly, we explore the potential experimental realization of the PTQRM based on superconducting circuits.
By including these aspects, we hope to offer a comprehensive perspective on the PTQRM Hamiltonian and its potential implications in the realm of quantum physics.

~

\subsection{Alternative representation of the PTQRM}

In this part, we further justify the $\mathcal{PT}$ symmetry of the PTQRM by drawing parallels with the conventional form of $\mathcal{PT}$-symmetric qubit in the existing literature \cite{Dogra_2021,Naghiloo2019,Oezdemir2019,ElGanainy2018}. 
Although the $\mathcal{PT}$ symmetry in the non-Hermitian qubit, as described by Eq.~(\ref{Hqubit}), might not seem obvious at first glance, a deeper examination reveals its underlying symmetry.
Here in Eq.~(\ref{Hqubit}), the superscript $z$ signifies the chosen basis where $\sigma_z$ is diagonalized.
Therefore, the two basis states of the qubit are $\ket{\pm} = \ket{\pm z}$.
In this framework, $\Delta$ represents the level splitting between the two states and $i\epsilon$ delineates the coupling between them. 

The direct physical interpretation of the imaginary coupling in $H_q^z$ might pose challenges.
To render it more intuitive, we can implement a unitary transformation generated by
\begin{equation}
    \quad U = \frac{1}{\sqrt{2}}\begin{pmatrix}
        1 & 1 \\
        1 & -1
    \end{pmatrix},
\end{equation}
with the effects
\begin{equation}
    \quad U^\dagger \sigma_x U = \sigma_z, \quad U^\dagger \sigma_z U = \sigma_x. 
\end{equation}
We then arrive at the new qubit Hamiltonian
\begin{equation}
    H_q^x = U^\dagger H_q^z U = \frac{\Delta}{2}\sigma_x + \frac{i\epsilon}{2}\sigma_z  = \frac{1}{2}\begin{pmatrix}
        i\epsilon & \Delta \\
        \Delta & - i \epsilon
    \end{pmatrix},
\end{equation}
where the superscript $x$ indicates our transition to the eigenbasis of $\sigma_x$. 
With $H_q^x$, the two states of the qubit are now $ \ket{\pm x}$.
The term $\Delta$ describes the mutual coupling of these states, while the imaginary diagonal terms $i\epsilon$ intuitively represent the gain for the excited state and the corresponding loss for the ground state.

Conventionally, a $\mathcal{PT}$-symmetric qubit is expressed in the form of $H_q^x$ due to the explicit physical interpretation of the imaginary terms \cite{Dogra_2021}. 
It is readily seen that $H_q^x$ remains invariant under the transformation
\begin{equation}
    (\mathcal{P}_x\mathcal{T})^\dagger H_q^x (\mathcal{P}_x\mathcal{T}) = H_q^x,
\end{equation}
where $\mathcal{P}_x = \sigma_x$ and $\mathcal{T}$ is complex conjugation. 

To ascertain the $\mathcal{PT}$ symmetry of our original qubit Hamiltonian $H_q^z$, we only need to revert to the $\sigma_z$ basis by doing the $U$ transformation again:
\begin{equation}
    H_q^z = U^\dagger H_q^x U. 
\end{equation}
This implies the parity operator adopts the form $\mathcal{P}_z = \sigma_z$, whereas the $\mathcal{T}$ operator remains the same. 
With this, the $\mathcal{PT}$ symmetry is evident from
\begin{equation}
    (\mathcal{P}_z\mathcal{T})^\dagger H_q^z (\mathcal{P}_z \mathcal{T}) = H_q^z.
\end{equation}

The PTQRM Hamiltonian, when expressed in this rotated representation, becomes
\begin{equation}\label{PTQRMx}
    \begin{aligned}
        H_\text{PTQRM}^x = &\frac{\Delta}{2}\sigma_x + \frac{i\epsilon}{2}\sigma_z + \omega a^\dagger a + g\sigma_z \left(a^\dagger + a \right) \\
        = & H_\text{QRM}^x + \frac{i\epsilon}{2}\sigma_z.
    \end{aligned}
\end{equation}
Notably, the Hermitian \(H_\text{QRM}^x\) aligns with the convention of the QRM within the context of superconducting circuits \cite{Irish_2005}.
Verification confirms that the Hamiltonian (\ref{PTQRMx}) has exactly the same eigenspectrum as the PTQRM. 
The corresponding eigenstates, along with other physical properties, remain congruent upon a rotation transformation. 

~

\subsection{The passive \texorpdfstring{$\mathcal{PT}$}{PT} symmetry}

In experimental realizations, the passive form of $\mathcal{PT}$ symmetry is often favored due to its practicality \cite{Naghiloo2019,Wang2021}. 
The passive $\mathcal{PT}$-symmetric qubit is described by 
\begin{equation}\label{pptqubit}
    \tilde{H}_q^x = \frac{\Delta}{2}\sigma_x + i\epsilon\sigma_+\sigma_- = H^x_q + \frac{i\epsilon}{2}\mathbb{I} .
\end{equation}
Upon examination, it becomes evident that the passive $\mathcal{PT}$-symmetric qubit $\tilde{H}_q^x$ possesses the same topology as its $\mathcal{PT}$-symmetric counterpart,  with the sole difference being a constant shift ${i\epsilon}/{2}$ in eigenvalues. 
Within the passive $\mathcal{PT}$ symmetry framework, only loss exists.
Nonetheless, it retains all the salient features inherent to standard $\mathcal{PT}$ symmetry, including EPs \cite{ElGanainy2018,Oezdemir2019}. 

Incorporating a cavity into the qubit as expressed in Eq.~(\ref{pptqubit}) naturally extends to the QRM under passive $\mathcal{PT}$ symmetry, represented by
\begin{equation}\label{passivePTQRM}
    \begin{aligned}
        \tilde{H}^x_\text{PTQRM} &=  \frac{\Delta}{2}\sigma_x + i\epsilon\sigma_+\sigma_- + \omega a^\dagger a + g\sigma_z \left(a^\dagger + a\right) \\
        & = H^x_\text{PTQRM} + \frac{i\epsilon}{2}\mathbb{I}.
    \end{aligned}
\end{equation}
The spectrum now has a constant shift $i\epsilon/2$ in energy, compared to both the PTQRM (\ref{PTQRM}) and its alternative representation \(H^x_\text{PTQRM}\). 
Moreover, the non-Hermitian dynamics generated by the Hamiltonian \(\tilde{H}^x_\text{PTQRM}\) are also identical to those of the PTQRM, upon undergoing the renormalization outlined in Eq.~(\ref{photonnumber}). 

Up to this point, we have delineated several forms of the PTQRM Hamiltonians as given in Eqs.~(\ref{PTQRM}), (\ref{PTQRMx}) and (\ref{passivePTQRM}). 
These distinct representations are interrelated as illustrated by the sequence
\begin{equation}
    H \xrightarrow{\text{rotation}} H^x_\text{PTQRM} \xrightarrow{\text{energy shift}} \tilde{H}^x_\text{PTQRM}.
\end{equation}
Crucially, their respective eigenspectra and dynamics coincide when considering energy offsets or appropriate wave function renormalizations.

~

\subsection{Lindblad master equation approach}

While closed quantum systems evolve according to the Schr\"odinger equation, open systems, those that interact with an external environment, often require a more intricate treatment.
In such scenarios, the Lindblad master equation (LME), a widely accepted formulation of the quantum master equation, becomes instrumental \cite{Manzano_2020}.  
This equation offers a framework to account for effects such as decoherence and dissipation that arise due to environmental interactions.
Here, we illustrate that the passive PTQRM (\ref{passivePTQRM}) can be perceived as the non-Hermitian effective Hamiltonian that emerges from the underlying dynamics captured by the LME. 

To better elucidate the link between the passive PTQRM and the dynamics driven by environmental effects, we consider a dissipative qubit coupled to a cavity.
The dissipative qubit may be described by a three-level system, and the corresponding basis states are $ \{ \ket{2}, \ket{1}, \ket{0} \} $ with decreasing energies \cite{Naghiloo2019}. 
We specify the decay rate, denoted by $\gamma$, from $\ket{1}$ to $\ket{0}$.
Such dynamics can be accurately captured through the LME, given by
\begin{equation}
    \frac{d\rho}{dt}=-i\left[H_\mathrm{QRM}^x,\rho\right]+\gamma\mathcal{D}\left[\ket{0}\bra{1}\right] \rho,
\end{equation}
where $H_\mathrm{QRM}^x$ is the standard QRM Hamiltonian in the rotated frame given in Eq.~(\ref{PTQRMx}). 
Here $\rho$ is the density matrix of the system, which encapsulates both the state information and coherence properties.
$\mathcal{D}[A]\rho\equiv A\rho A^\dagger-\frac{1}{2}(A^\dagger A \rho+\rho A^\dagger A)$ defines the Lindblad superoperator. 

We then expand the superoperator and rearrange the terms to obtain
\begin{equation}\label{LMEwithjump}
    \begin{aligned}
        \frac{d\rho}{dt}=&-i \left(H_\mathrm{QRM}^x - i\frac{\gamma}{2} \ket{1}\bra{1}\right)\rho \\
        &+i\rho \left( H_\mathrm{QRM}^x + i\frac{\gamma}{2}\ket{1}\bra{1}\right)+ \gamma \ket{0}\rho_{11}\bra{0}.
    \end{aligned}
\end{equation}
Upon confining our attention to the subspace $\mathrm{span}\{\ket{1},\ket{2}\}$ and with the identification $\gamma=2\epsilon$, the governing dynamics of the system, in terms of the LME, morphs into
\begin{equation} \label{MasterEquation}
     \frac{d\rho}{dt}= -i\left(\tilde{H}\rho-\rho \tilde{H}^\dagger\right) + 2\epsilon\ket{0} \rho_{11}\bra{0}.
\end{equation}
Here, the effective Hamiltonian $\tilde{H}=\tilde{H}^x_\text{PTQRM}$ is synonymous with our passive PTQRM \eqref{passivePTQRM} in the rotated frame, which is equivalent to our standard PTQRM Hamiltonian \eqref{PTQRM}. 
The last term in Eq.~(\ref{LMEwithjump}), known as the quantum jump operator, incorporates the transitions from the state space of the effective Hamiltonian to the larger Hilbert space of the environment. 

Upon neglecting the quantum jump term, the LME reduces to the von Neumann equation, which is formally equivalent to the TDSE of the effective Hamiltonian \cite{Manzano_2020}. 
Consequently, the system's dynamics are again governed by the effective non-Hermitian Hamiltonian \(\tilde{H}^x_\text{PTQRM}\), which parallels the PTQRM Hamiltonian (\ref{PTQRM}).
The distinction between LME and TDSE arises from the influence of the quantum jump term.

Experimentally, when one solely focuses on the effective Hamiltonian, a post-selection scheme can be applied.
This technique eliminates the effects of quantum jump terms \cite{Naghiloo2019}, rendering the system dynamics to be wholly governed by the non-Hermitian Hamiltonian.

~

\subsection{Potential physical realizations}

\begin{figure}[t]
	\includegraphics[width=\linewidth]{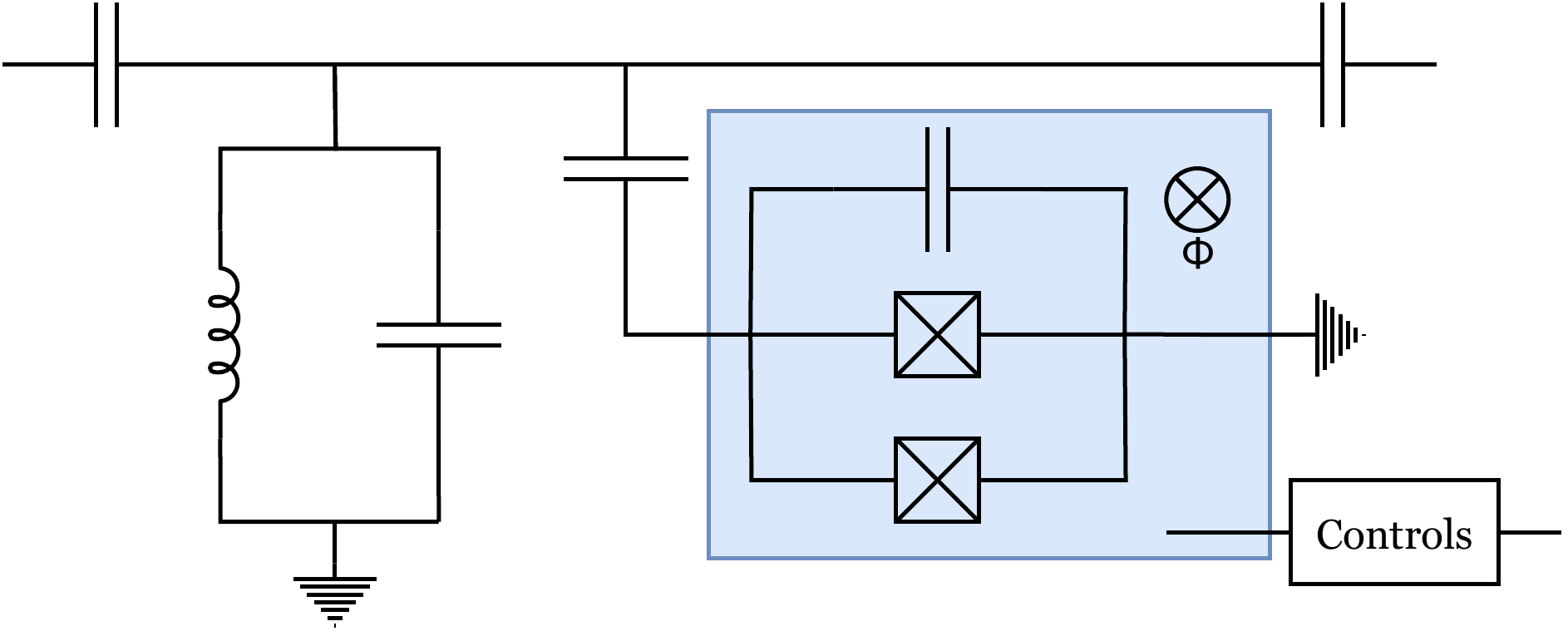}
	\caption{Possible circuit QED realization, the transmon qubit circuit is embedded in a cavity (blue box in this diagram) with controls, for driving and read-out, explained in Ref.~\cite{Naghiloo2019} in detail. The lowest three energy levels are used for constructing a $\mathcal{PT}$-symmetric qubit, which is consistent with the master equation approach. Decay rates between these states are controlled by the magnetic flux $\Phi$.}
	\label{PTQRMcQED}
\end{figure}

From the perspective of circuit quantum electrodynamics (QED), we could envision a synthesis by merging the realization of QRM in Ref.~\cite{Yoshihara_2016} and the $\mathcal{PT}$-symmetric qubit in Ref.~\cite{Naghiloo2019}. 
The resulting superconducting circuit is depicted in Fig.~\ref{PTQRMcQED}, where a transmon qubit contained in a cavity is coupled to an LC resonator. 
Drawing from the transmon circuit's lowest three energy levels, we proceed with the derivation of the master equation for the passive PTQRM: the ground state functions as a sink for the excited (qubit) states. 
It was shown in \cite{Naghiloo2019} that, by modulating the magnetic flux $\Phi$ through the transmon circuit, we can control the two decay rates such that $\ket{1}$ to $\ket{0}$ is much faster than $\ket{2}$ to $\ket{1}$.
Given this differential in decay rates, we can judiciously overlook the latter transition, leading to a characterization of this system via our master equation \eqref{MasterEquation} in the absence of the quantum jump term. 

In addition to the circuit QED platform, trapped ion experiments, employing the $ {}^{40}\mathrm{Ca}^+$ ion \cite{Wang2021}, also offer a prospective avenue for realizing our system, specifically by incorporating the lasing mode.
A noteworthy investigation has recently delved into this avenue, adding an auxiliary 1-photon state, thus revealing a more intricate landscape of EP structures \cite{Kim2023}. 

Another possible direction is considering the substitution of the traditional qubit in QRM experiments \cite{Cai2021} with $\mathcal{PT}$-symmetric ones.
This potential pivot points to an interesting direction in the field.

~

\section{Summary}

In this work, we explored the intricacies of the $\mathcal{PT}$-symmetric quantum Rabi model (PTQRM) and formulated an analytical solution via the adiabatic approximation (AA). 

Our exploration of static properties revealed multiple exceptional points (EPs) within the spectrum. 
Intriguingly, these EPs vanish and revive depending on the coupling strength. 
A significant observation was the identification of unique points where distinct eigenstates stand orthogonal -- a characteristic absent in typical non-Hermitian systems. 
In particular, these orthogonal points coincide with the level crossing points of the Hermitian QRM.

We conducted a comprehensive examination of non-Hermitian dynamics in the strong coupling (SC) and ultrastrong coupling (USC) regimes. 
In the SC regime, the intricate interplay between Hermitian and non-Hermitian couplings manifests strikingly distinct dynamical phenomena on either side of the EPs. 
In contrast, within the USC regime, our investigation revealed a three-stage evolution pattern in the cavity photon number: oscillation around the initial state, convergence to nearby eigenstates with large imaginary energies, and divergence to infinite-photon states.

This work broadens the theoretical horizon of $\mathcal{PT}$ symmetry, encompassing the comprehensive quantum light-matter interaction domain.
Furthermore, it offers insights that have the potential to be extrapolated across a wide spectrum of quantum optical systems \cite{Li2021a, Xie_2014, Xie_2019}. 
We are optimistic about the practical implementation of the PTQRM across various experimental platforms, such as circuit QED \cite{Naghiloo2019} and trapped ions \cite{Wang2021}.

~

\begin{acknowledgments}
This work is supported by the National Natural Science Foundation of China Grant No. 12205383, and the Natural Science Foundation of Changsha Grant No. kq2202082, and Australian Research Council Grant No. DP210102243.

X. Lu, H. Li, and J.-K. Shi contributed equally to this work. 
\end{acknowledgments}

\appendix

~

\section{\texorpdfstring{$\mathcal{PT}$}{PT} symmetry}\label{AppendixPT}

\begin{figure}[t]
    \centering
    \includegraphics[width=\linewidth]{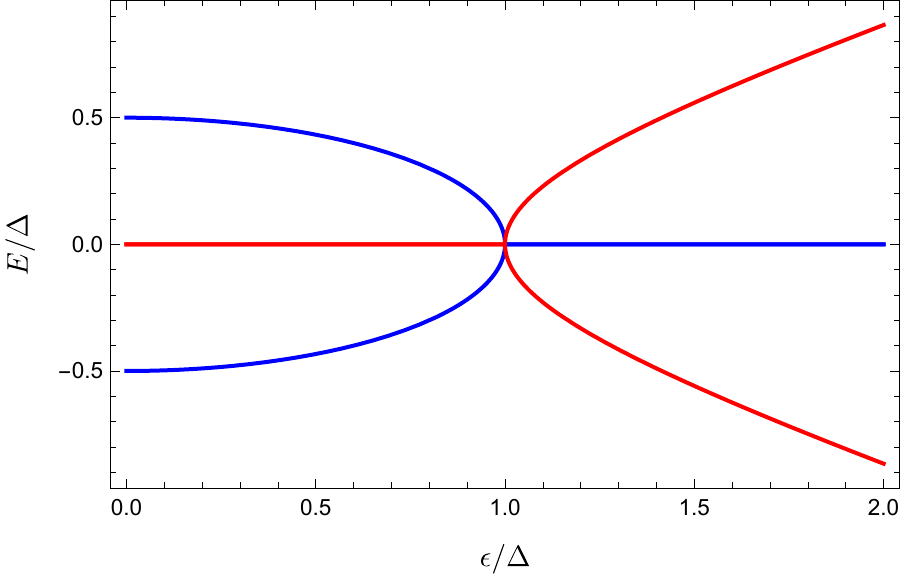}
    \caption{The energy eigenvalues of the $\mathcal{PT}$-symmetric qubit $H^z_q$.
    The real part is in blue lines and the imaginary part is in red lines. 
    }
    \label{PTqubitSpectrum}
\end{figure}

It is straightforward to calculate the eigenstates and eigenvalues of the non-Hermitian qubit defined in the main text, i.e.,
\begin{equation}
    H_q^z = \frac{\Delta}{2}\sigma_z + \frac{i\epsilon}{2}\sigma_x.
\end{equation}
The eigensystem can be divided into two cases, depending on the values of $\Delta$ and $\epsilon$, as explained below. 

\textit{Case I}: $\Delta \ge \epsilon$. 
The eigenvalues and corresponding eigenstates are
\begin{equation}
    \begin{aligned}
        E_\pm^z &= \pm\frac{1}{2}\sqrt{\Delta^2-\epsilon^2}, \\ 
    \ket{\psi^z_\pm} &= \begin{pmatrix}
        \dfrac{-i\Delta \mp i \sqrt{\Delta^2 - \epsilon^2}}{\epsilon} \\ \vspace{-.5em}\\
        1        
    \end{pmatrix}.
    \end{aligned}
\end{equation}
It is easy to verify that $\ket{\psi^z_\pm}$ are also the eigenstates of the combined $\mathcal{PT}$ operator, namely
\begin{equation}
    \mathcal{PT} \ket{\psi^z_\pm} = - \ket{\psi^z_\pm}.
\end{equation}
Therefore, the system is in the $\mathcal{PT}$-symmetric phase. 

\textit{Case II}: $\Delta < \epsilon$. 
The eigenvalues and corresponding eigenstates are
\begin{equation}
    \begin{aligned}
        E_\pm^z &= \pm \frac{i}{2}\sqrt{\epsilon^2 - \Delta^2}, 
        \\
        \ket{\psi^z_\pm} &= \begin{pmatrix}
            \dfrac{-i\Delta \pm  \sqrt{\epsilon^2 - \Delta^2}}{\epsilon}\\ \vspace{-.5em} \\ 1 
        \end{pmatrix} ,
    \end{aligned}
\end{equation}
which now yields
\begin{equation}\label{PTbroken}
    \mathcal{PT} \ket{\psi^z_\pm} = \begin{pmatrix}
        \dfrac{i\Delta \pm  \sqrt{\epsilon^2 - \Delta^2}}{\epsilon}\\ \vspace{-.5em} \\ -1 
    \end{pmatrix}=-\ket{\psi_\mp^z}
        \ne c \ket{\psi^z_\pm},
\end{equation}
where $c \in \mathbb{C}$.
This case is termed as $\mathcal{PT}$-broken because the wave functions do not have $\mathcal{PT}$ symmetry even though the Hamiltonian still commutes with the combined $\mathcal{PT}$ operator. 
An interesting fact is that the combined $\mathcal{PT}$ operator flips the two eigenstates in the $\mathcal{PT}$-broken case, as shown in Eq.~(\ref{PTbroken}).
The energy spectrum discussed above is depicted in Fig.~\ref{PTqubitSpectrum}, with the real part and imaginary part presented in blue and red lines, respectively. 
The transition from real to complex eigenvalues emerges at the exceptional point $\epsilon/\Delta=1$. 

Regarding the $\mathcal{PT}$ symmetry of the light field, we recall the general effects of parity and time operators $ \mathcal{P}$ and $ \mathcal{T} $ described as 
\begin{equation}
    \begin{split}
        \mathcal{P}: i\rightarrow i,\quad x\rightarrow -x,\quad p\rightarrow -p \\
    \mathcal{T}: i\rightarrow -i,\quad x \rightarrow x, \quad p\rightarrow -p
    \end{split}
\end{equation}
It follows that by applying the $ \mathcal{PT} $ transformation to the bosonic creation and annihilation operators, we have
\begin{equation}
    a\rightarrow -a,\quad a^\dagger \rightarrow -a^\dagger,
\end{equation}
which is equivalent to the bosonic parity transformation \cite{Li2021a}.

\end{document}